\documentclass[letterpaper, 10 pt, conference]{ieeeconf}  

\IEEEoverridecommandlockouts                              

\title{\LARGE \bf
Simultaneous Controller and Lyapunov Function Design for Constrained Nonlinear Systems 
}

\author{Reza Lavaei$^{1}$, Leila Bridgeman$^{1}$
\thanks{$^{1}$Reza Lavaei and Leila Bridgeman are with the Department of Mechanical Engineering and Materials Science at Duke University, Durham NC, USA (email:  reza.lavaei@duke.edu; leila.bridgeman@duke.edu), corresponding author: Reza Lavaei}
}

\usepackage[utf8]{inputenc}
\usepackage{graphicx}
\usepackage{amsmath}
\usepackage{xfrac}
\DeclareMathOperator*{\argmin}{argmin}

\usepackage{gensymb}

\usepackage[nolist]{acronym}

\usepackage{braket}

\usepackage{physics}

\usepackage{amssymb,mathtools}
\usepackage{subcaption}

\usepackage[font=small,labelfont=bf]{caption}
\usepackage[font=small,labelfont=bf]{subcaption}

\usepackage{hyperref}

\usepackage{amsfonts}
\usepackage{nicefrac}

\usepackage[table,xcdraw,dvipsnames]{xcolor}

\usepackage{algorithm} 
\usepackage{algpseudocode}
\usepackage{cite}

\usepackage[final]{pdfpages}

\usepackage{pgfplots}
\pgfplotsset{compat=newest}
\pgfplotsset{plot coordinates/math parser=false} 

\usepackage{tabu}

\newtheorem{definition}{Definition}

\newtheorem{corollary}{Corollary}
\newtheorem{theorem}{Theorem}

\newtheorem{remark}{Remark}
\newtheorem{initialization}{Initialization}

\providecommand{\openbox}{\leavevmode
  \hbox to.77778em{%
  \hfil\vrule
  \vbox to.675em{\hrule width.6em\vfil\hrule}%
  \vrule\hfil}}
\makeatletter
\DeclareRobustCommand{\qed}{%
  \ifmmode
    \eqno \def\@badmath{$$}
    \let\eqno\relax \let\leqno\relax \let\veqno\relax
    \hbox{\openbox}%
  \else
    \leavevmode\unskip\penalty9999 \hbox{}\nobreak\hfill
    \quad\hbox{\openbox}%
  \fi
}
\makeatother
\newcommand{\mt}{m_{\mathcal{T}}}
\newcommand{\Et}{\mathbb{E}_\mathcal{T}}
\newcommand{\T}{\mathcal{T}}
\newcommand{\X}{\mathcal{X}}
\newcommand{\IntSet}{\mathbb{Z}}
\newcommand{\R}{\mathbb{R}}
\newcommand{\A}{\mathcal{A}}

\begin{document}

\maketitle
\thispagestyle{empty}
\pagestyle{empty}

\begin{abstract}
This paper presents a method to stabilize state and input constrained nonlinear systems using an offline optimization on variable triangulations of the set of admissible states. For control-affine systems, by choosing a \ac{CPA} controller structure, the non-convex optimization is formulated as iterative \ac{SDP}, which can be solved efficiently using available software. The method has very general assumptions on the system's dynamics and constraints. Unlike similar existing methods, it avoids finding terminal invariant sets, solving non-convex optimizations, and does not rely on knowing a \ac{CLF}, as it finds a \ac{CPA} Lyapunov function explicitly. The method enforces a desired upper-bound on the decay rate of the state norm and finds the exact region of attraction. Thus, it can be also viewed as a systematic approach for finding Lipschitz \ac{CLF}s in state and input constrained control-affine systems. Using the \ac{CLF}, a minimum norm controller is also formulated by quadratic programming for online application. 
\end{abstract}
\section{INTRODUCTION}
The dynamics of most systems, like autonomous vehicles \cite{Ames2016}, robotics \cite{robot2020}, and chemical processes \cite{chemProccess2013}, is constrained. Inputs are constrained by actuation capability, and state constraints are either imposed by physical limitations or safety considerations. Depending on the dynamics and constraints, different methods exist to find a controller that ensures existence of a Lyapunov function, and thus Lyapunov stability. Two major approaches are \ac{MPC} \cite{nonMPC1998} and Lyapunov-based methods \cite{Ames2016}. This work presents a method that for control-affine systems improves on \ac{MPC} by avoiding non-convex optimizations, and on Lyapunov-based methods by not requiring a known \ac{CLF}.

Most nonlinear MPC formulations depend on careful choices of terminal `ingredients' that consist a set, a cost function, and a stabilizing controller \cite{nonMPC1998}. Other \ac{MPC} approaches either ensure existence of these ingredients implicitly \cite[Ch.\;2.3]{nonMPCeconomic2018}, or circumvent them using Lyapunov-based \ac{MPC} \cite{nonMPC-Lyap2005,nonMPC-Lyap2006} if a \ac{CLF} is known. These methods often rely on solving a non-convex optimization online. Using numerical methods to solve them not only does not guarantee finding the global solution, but is computationally taxing, making their efficient implementation a question of ongoing research \cite{nonMPCefficient2009,nonMPCefficient2020}. To avoid online optimization, explicit nonlinear \ac{MPC} finds the controller offline \cite{nonMPCExTube2006,nonMPCexplicit2012}. However, solving a highly nonlinear optimization on \textit{a priori} unknown polyhedral partitions remains difficult.

Lyapunov-based methods rely on Lyapunov-like functions, such as \ac{CLF}s and control barrier functions, to ensure stability of control-affine systems. While barrier functions \cite{CBF2007,CBF2016,CBF2018} impose state constraints by ensuring positive invariance of a subset of admissible states, satisfying input constraints needs the conditions on the Lyapunov-like functions' time derivatives to hold for admissible inputs \cite{uConCLF1995,Ames2016}. If such functions are known, online \ac{QP} can be used to find a minimum-norm controller that not only ensures safety and stability, but can also prioritize safety if needed \cite{Ames2016}. However, these methods require Lyapunov-like functions, which are not trivial to find. 

This paper presents a method to stabilize state and input constrained nonlinear systems via an offline optimization on variable trianglations of admissible states that refines simplexes if needed. Since it finds the corresponding \ac{CPA} Lyapunov function explicitly, the exact region of attraction and an upper-bound on the decay rate of the state norm are provided. By choosing a \ac{CPA} state feedback controller structure, the nonlinear optimization is solved iteratively using SDPs for control-affine systems. In this case, the corresponding Lipschitz \ac{CLF} is used to formulate a minimum-norm controller by \ac{QP}. In seeking both the controller and the Lyapunov function offline, this method is similar to \cite{ExpMPCLike2017}, but it is not limited to polynomial systems. Like \cite{nonMPCExTube2006}, the method depends on refining elements in a subset of the state space, but it avoids solving the highly nonlinear optimization. The method builds upon the analysis technique of \cite{gieslRevCPA2013} that implements \ac{CPA} Lyapunov functions. For control-affine systems, it improves \cite{Doban}, which is also based on \cite{gieslRevCPA2013}, by not requiring a known \ac{CLF}, and removing the need for \textit{a priori} constraints on the controller's gradient.

\section{Preliminaries}
\textbf{Notation.} The interior, boundary, and closure of $\Omega\in\R^n$ are denoted by $\Omega\degree$, $\partial\Omega$, and $\bar{\Omega}$, respectively. The set of real-valued functions with $r$ times continuously differentiable partial derivatives over their domain is denoted by $\mathbb{C}^r$. The $i$-th element of a vector $x$ is denoted by $x^{(i)}$. The element in the $i$-th row and $j$-th column of a matrix $G$ is denoted by $G^{(i,j)}$. The preimage of a function $f$ with respect to a subset $\Omega$ of its codomain is defined by $f^{-1}(\Omega)=\{x\mid f(x) \in \Omega \}$. The transpose and Euclidean norm of $x\in\R^n$ are denoted by $x^\intercal$ and $||x||$, respectively. The set of all subsets $\Omega\subset\R^n$ satisfying i) $\Omega$ is compact, ii) $\Omega\degree$ is a connected open neighborhood of the origin, and iii) $\Omega=\overline{\Omega\degree}$ is denoted by $\mathfrak{R}^n$. The vector of ones in $\R^n$ is denoted by $1_n$. 

The exponential stability of an autonomous system's equilibrium point can be verified by constructing a Lipschitz Lyapunov function on a triangulated subset of $\R^n$ \cite{gieslRevCPA2013}. The required definitions are given next.

\begin{definition}[Affine independence{\cite{gieslRevCPA2013}}] \label{def:affDepVecs}
A collection of vectors $\{x_0,\ldots,x_n\}$ in $\R^n$ is called affinely independent if $x_1-x_0,\ldots,x_n-x_0$ are linearly independent. \qed
\end{definition}

\begin{definition}[$n$-simplex {\cite{gieslRevCPA2013}}] \label{def:simplex}
An $n$-simplex is the convex combination of $n+1$ affinely independent vectors in $\R^n$, denoted $\sigma{=}\textrm{co}(\{x_j\}_{j=0}^n)$, where $x_j$'s are called vertices. \qed
\end{definition}

\noindent In this paper, simplex always refers to $n$-simplex. By abuse of notation, $\T$ will refer to both a collection of simplexes and the set of points in all the simplexes of the collection. 

\begin{definition} [Triangulation {\cite{gieslRevCPA2013}}] \label{def:triangulation}
A set $\T\in\mathfrak{R}^n$ is called a triangulation if it is a finite collection of $\mt$ simplexes, denoted $\T=\{\sigma_i\}_{i=1}^{\mt}$, and the intersection of any of the two simplexes in $\T$ is either a face or the empty set. 

The following two conventions are used throughout this paper for triangulations and their simplexes. Let $\T=\{\sigma_i\}_{i=1}^n$. If $0\in\sigma_i$, then $0$ is a vertex of $\sigma_i$. Further, let $\{x_{i,j}\}_{j=0}^n$ be the vertices of simplex $\sigma_i$. Then $\sigma_i$ is represented by $\sigma_i=\textrm{co}(\{x_{i,j}\}_{j=0}^n)$. The choice of $x_{i,0}$ in $\sigma_i$ is arbitrary unless $0\in\sigma_i$, where $x_{i,0}$ is selected as $0$. The vertices in the triangulation $\T$ is denoted by $\Et$. \qed
\end{definition}

\begin{definition} [CPA interpolation \cite{gieslRevCPA2013}] \label{def:CPAfunction}
Consider a triangulation $\T{=}\{\sigma_i\}_{i=1}^{\mt}$, and a set $\mathbf{W}{=}\left\{ W_x \right\}_{ x\in \Et } {\subset} \R$. The unique, CPA interpolation of $\textbf{W}$ on $\T$, denoted $W:\T{\rightarrow}\R$, is affine on each $\sigma_i{\in}\T$ and satisfies $W(x){=}W_x$, $\forall x{\in}\Et$. \qed
\end{definition}

\begin{remark}[{\!\!\cite[Rem.\;9]{gieslRevCPA2013}}] \label{rem:nablaLinear}
    Given $\T=\{\sigma_i\}_{i=1}^{\mt}$ and $\mathbf{W}$, the CPA interpolation assigns a unique affine function $W(x)=x^\intercal\nabla{W}_i+\omega_i$ to each $\sigma_i\in\T$. The $\nabla{W}_i$ is linear in the elements of $\mathbf{W}$ and can be computed as follows. Let $\sigma_i=\textrm{co}(\{x_{i,j}\}_{j=0}^n)$, and $X_i\in\R^{n\times n}$ be a matrix that has $x_{i,j}-x_{i,0}$ as its $j$-th row. Since the elements of $\{x_{i,j}\}_{j=0}^n$ are affinely independent, $X_i$ is invertible. Each $x_{i,j}$ is an element of $\Et$, so it has a corresponding element in $\mathbf{W}$, denote $W_{x_{i,j}}$. Let $\bar{W}_i\in\R^n$ be a vector that has $W_{x_{i,j}}-W_{x_{i,0}}$ as its $j$-th element. Then, ${\nabla W}_i = X^{-1}_i \bar{W}_i$. \qed
\end{remark}

The following theorem from \cite{gieslRevCPA2013} bounds the time derivative of a CPA function above on a simplex using its values at the vertices of that simplex using Taylor's theorem. 

\begin{theorem}[{\!\!\cite{gieslRevCPA2013}}] \label{thm:gieseldWBound}
Consider the system
\begin{equation} \label{eq:AutSystem}
 \dot{x} = g(x), \;\; x\in \X \in \mathfrak{R}^n, 
 \end{equation}
where $g:\R^n\rightarrow\R^n$ is in $\mathbb{C}^2$. Let $\T=\{\sigma_i\}_{i=1}^{\mt} \subseteq \X$ be a triangulation, and $W(x):\T\rightarrow\R$ be the CPA interpolation of a set $\mathbf{W}=\{W_x\}_{x\in{\Et}}$. Consider a point $x\in\T\degree$. The Dini derivative of $W$ at $x$ is defined as $D^+W(x) = \textrm{lim\,sup}_{h\rightarrow0^+}\sfrac{(W(x+hg(x))-W(x))}{h}$, which equals $\dot{W}(x)$ where $W\in\mathbb{C}^1$. For an arbitrary $x\in\T\degree$, there exists a $\sigma_i=\{x_{i,j}\}_{j=0}^n\in\T$ so that for small enough $h>0$, $\textrm{co}(x,x+hg(x))\subset\sigma_i$. Let $0\leq\alpha_j\leq 1$, where $j\in\IntSet_0^n$ and $\sum_{j=0}^n\alpha_j=1$, be the unique set of coefficients satisfying $x=\sum_{j=0}^n\alpha_ix_{i,j}$. Then
\begin{equation} \label{eq:gieselInequality}
    D^+W(x) \leq \sum_{j=0}^n \alpha_j \left( g(x_{i,j})^\intercal\nabla{W}_i + c_{i,j}\beta_i 1_n^\intercal l_i \right),
\end{equation}
where $l_i\in\mathbb{R}^n$ satisfies $l_i\geq|\nabla{W}_i|$, and 
\begin{flalign} 
    &\beta_i \geq \max_{p,q,r\in\IntSet_1^n} \max_{\xi\in\sigma_i} \left| \left. \sfrac{\partial^2 g^{(p)}}{\partial x^{(q)}\partial x^{(r)}} \right|_{x=\xi} \right|, \textrm{ and} \label{eq:beta} \\
    &c_{i,j}{=}\frac{n}{2} ||x_{i,j} {-} x_{i,0}|| (\max_{k\in\IntSet_1^n} ||x_{i,k}{-}x_{i,0}|| {+} ||x_{i,j}{-}x_{i,0}||). \nonumber
\end{flalign}
\qed
\end{theorem}

Note that in \eqref{eq:beta}, $\beta_i$ bounds the largest absolute value of the elements of the Hessian of $g(x)$ on $\sigma_i$ above.
\section{Control Design}
Using Theorem\;\ref{thm:gieseldWBound}, the exponential stability of of the equilibrium can be verified by constructing a CPA Lyapunov function formulated as a linear feasibility program \cite{gieslRevCPA2013}. Here, the goal is to turn the analysis method of \cite{gieslRevCPA2013} into a design method for state and input constrained control systems by finding a state-feedback controller that makes the origin exponentially stable. Choosing a parameterized controller structure, the search for parameters can be formulated as a non-convex optimization since the CPA Lyapunov function is also unknown. First, a stability theorem and piecewise twice continuous differentiability on a triangulation are defined and then, the optimization is formulated. The following theorem, improves on \cite[Def\;2,\;Rem\;5]{gieslRevCPA2013} by bounding the convergence rate of $||x(t)||$ above.

\begin{theorem} \label{thm:myExpoStability}
The origin in \eqref{eq:AutSystem}, where $g:\Omega\rightarrow\R^n$ is a Lipschitz map, $\Omega\in\mathfrak{R}^n$, and $g(0)=0$, is exponentially stable if there exists a Lipschitz function $V:\Omega\rightarrow\R^n$ and constants $a,b_1,b_2>0$ satisfying $V(0)=0$, and
\begin{subequations} \label{eq:myExpo}
    \begin{align}
         b_1||x||^a &\leq V(x), \quad \forall x\in\Omega, \textrm{ and} \label{eq:myExpoBound}\\
 D^+V(x) & \leq -b_2 V(x), \quad \forall x\in\Omega\degree \backslash \{0\}. \label{eq:myExpoDecrease}
    \end{align}
\end{subequations}
Further, let $\A=V^{-1}([0,r])\subseteq\Omega$ be in $\mathfrak{R}^n$ for some $r>0$. Then, $||x(t)||\leq\sqrt[^a]{r/b_1}e^{(-b_2/a)(t-t_0)}$, $\forall x(t_0)\in\A\degree$. \qed
\end{theorem}

\begin{proof}
Since $\A\in\mathfrak{R}^n$, $x(t_0)\in\A\degree$ implies $V(x(t)) \leq r$ and $x(t)\in\A\degree$ holds for all $t\geq t_0$. Using the Comparison Lemma \cite[Lem 3.4]{khalil}, $V(x(t))\leq V(x(t_0))e^{-b_2(t-t_0)}$ for all $t\geq t_0$. So, $||x(t)||\leq \sqrt[^a]{V(x(0))/b_1} e^{-(b_2/a)(t-t_0)}$, and therefore $||x(t)||\leq\sqrt[^a]{r/b_1}e^{-(b_2/a)(t-t_0)}$.
\end{proof}

\begin{definition} \label{def:C2functionsPiecewise}
A continuous function $g(x)\in\R^n$ is piecewise in $\mathbb{C}^2$ on a triangulation $\T=\{\sigma_i\}_{i=1}^{\mt}$, denoted $g\in\mathbb{C}^2(\T)$, if it is in $\mathbb{C}^2$ on $\sigma_i$ for all $i\in\IntSet_1^{\mt}$. \qed
\end{definition}

From now on, in case $\xi\in\sigma_i$, $\left.\sfrac{\partial^2 g^{(p)}}{\partial x^{(q)} \partial x^{(r)}}\right|_{x=\xi}$ for any vector function $g(x)$ and $p,q,r\in\IntSet_{i=1}^n$ means that the derivatives at the point $x=\xi$ are evaluated in those directions $y\in\R^n$ in which $\textrm{co}(x,x+hy)\subset \sigma_i$ as $h\rightarrow 0$.

\begin{theorem} \label{thm:genOpt}
Consider the system
\begin{align} \label{eq:controlSystem}
    \dot{x} = g(x,u), \; x\in\X\in\mathfrak{R}^n, \; u\in\mathcal{U}\in\mathfrak{R}^m, \; g(0,0)=0.
\end{align}
Given a triangulation $\T =\{\sigma_i\}_{i=1}^{\mt}$, where $\T\subseteq\X$, suppose that a class of Lipschitz controllers $\mathcal{F}=\{u(\cdot,\boldsymbol{\lambda})\}$ parameterized by $\boldsymbol{\lambda}$ is chosen so that $u(0,\lambda)=0$, and $g_{\boldsymbol{\lambda}}(\cdot) \coloneqq g(\cdot,u(\cdot,\boldsymbol{\lambda}))$ is Lipschitz on $\T$, and both $u(\cdot,\boldsymbol{\lambda})$, $g_{\boldsymbol{\lambda}}(\cdot)\in\mathbb{C}^2(\T)$, and $u(\cdot,\boldsymbol{\lambda})\in\mathcal{U}$ for $\forall x\in\Et$ implies $u(\cdot,\boldsymbol{\lambda})\in\mathcal{U}$ for $\forall x\in\T$, and $\mathcal{F}$ has an admissible element. Consider the following nonlinear program.
\begin{subequations} \label{eq:genOpt} 
    \begin{alignat}{2}
        [\mathbf{V}^\ast,\; &\mathbf{L}^\ast,\; \boldsymbol{\lambda}^\ast,\; a^\ast,\; \mathbf{b}^\ast] = && \argmin_{\mathbf{V},\; \mathbf{L},\; \boldsymbol{\lambda},\; a,\; \mathbf{b}} \;\; \hat{J}(\mathbf{V}, \mathbf{L}, \boldsymbol{\lambda}, a, \mathbf{b})  \nonumber \\
        \textrm{s.t.}\;\;& V_0 = 0, \;\; a,b_1 > 0, && \label{eq:V0b1Constraint} \\
        & b_1||x||^a \leq V_x, &&\forall x\in\Et\backslash\{0\}, \label{eq:myVconstraint} \\
        & |{\nabla V}_i| \leq l_i, &&\forall i\in\IntSet_1^{\mt}, \label{eq:nablaConstraint} \\
        & u(x_{i,j},\boldsymbol{\lambda})\in\mathcal{U}, &&\forall i\in\IntSet_1^{\mt}, \; \forall j\in\IntSet_0^n, \label{eq:uConstraintGeneral} \\
        & D^+_{i,j}V \leq -b_2 V_{x_{i,j}}, \;\; &&\forall i\in\IntSet_1^{\mt}, \; \forall j\in\IntSet_0^n, \label{eq:myDv}
        \end{alignat}
\end{subequations}

\noindent where $D^+_{i,j}V=g_{\boldsymbol{\lambda}}(x_{i,j})^\intercal {\nabla V}_i + c_{i,j}\beta_i 1_n^\intercal l_i$, and $\mathbf{V}=\{V_x\}_{x\in\Et}\subset\R$ and $\mathbf{L}=\{l_i\}_{i=1}^{\mt}\subset\R^n$, and $\mathbf{b}=\{b_1,b_2\}\subset\R$, and $\hat{J}$ is a cost function, and for $u(\cdot,\boldsymbol{\lambda})$ satisfying \eqref{eq:uConstraintGeneral},
\begin{flalign}\label{eq:betaAndc}
    &\beta_i \geq \max_{p,q,r\in\IntSet_1^n} \max_{\xi\in\sigma_i} \left| \left. \sfrac{\partial^2 g^{(p)}_{\boldsymbol{\lambda}}}{\partial x^{(q)}\partial x^{(r)}} \right|_{x=\xi} \right|, \textrm{ and} & \\
    &c_{i,j}{=} \frac{n}{2} ||x_{i,j} {-} x_{i,0}|| (\max_{k\in\IntSet_1^n} ||x_{i,k}{-}x_{i,0}|| {+} ||x_{i,j}{-}x_{i,0}||).& \nonumber
\end{flalign}

\noindent The optimization \eqref{eq:genOpt} is feasible. If $b_2^\ast > 0$ in \eqref{eq:genOpt}, then the CPA function $V^\ast:\T\rightarrow\R$ constructed from the elements of $\textbf{V}^\ast$ is a Lyapunov function of $\dot{x}=g_{\boldsymbol{\lambda}^\ast}(x)$. Let $\A = V^{\ast^{-1}}([0,r])\subseteq\T$ be in $\mathfrak{R}^n$ for some $r>0$. Then $x=0$ is locally exponentially stable for $\dot{x}=g_{\boldsymbol{\lambda}^\ast}(x)$ with $||x(t)||\leq\sqrt[^{a^\ast}]{r/b_1^\ast}e^{-(b_2^\ast/a^\ast)(t-t_0)}$ if $x(t_0)\in\A\degree$. \qed
\end{theorem}

\begin{proof}
To see that \eqref{eq:genOpt} is feasible, note that $V_x=b_1||x||^a$ with any $a,b_1>0$ satisfies \eqref{eq:myVconstraint} and can be used to compute a feasible solution $l_i=|\nabla{V}_i|$ for \eqref{eq:nablaConstraint} using Remark\;\ref{rem:nablaLinear}. By assumption, a feasible $\boldsymbol{\lambda}$ exists satisfying \eqref{eq:uConstraintGeneral}. Using these feasible values, finite $\beta_i$ satisfying \eqref{eq:betaAndc} can be chosen and $g_{\boldsymbol{\lambda}}(\cdot)$ is always finite because $g_{\boldsymbol{\lambda}}(\cdot)\in\mathbb{C}^2(\T)$. Likewise, $c_{i,j}$ is finite because each $\sigma_i$ is compact, making the left-hand side of \eqref{eq:myDv} finite for each $i\in\IntSet_1^{\mt}$ and $j\in\IntSet_0^n$. Note that if $x_{i,j}=0$, then $g_{\boldsymbol{\lambda}}(x_{i,j})=0$ and by convention, $j=0$, so $c_{i,j}=0$, making $D^+_{i,j}V=0$, making any $b_2$ feasible. Thus, there exists $b_2{\in}\R$ that satisfies \eqref{eq:myDv} for all $i{\in}\IntSet_1^{\mt}$ and $j{\in}\IntSet_0^n$. 

The remainder of the proof is devoted to showing that $V^\ast$ for the closed-loop system, $\dot{x}=g_{\boldsymbol{\lambda}}(x)$, verifies Theorem\;\ref{thm:myExpoStability} because by assumption,  \eqref{eq:uConstraintGeneral} implies $u(x,\boldsymbol{\lambda})\in\mathcal{U}$ for all $x\in\T$. Let $\Omega:=\T$ in Theorem\;\ref{thm:myExpoStability}. Constraints \eqref{eq:V0b1Constraint}--\eqref{eq:myVconstraint} ensure $V^\ast(0)=0$ and \eqref{eq:myExpoBound} since $V^\ast$ is a CPA function. It remains to show that \eqref{eq:nablaConstraint} and \eqref{eq:myDv} verify \eqref{eq:myExpoDecrease}. For simplicity, let $g(x)=g(x,u(\boldsymbol{\lambda},x))$. The assumptions of Theorem\;\ref{thm:gieseldWBound} with $W\coloneqq V$ are verified by \eqref{eq:nablaConstraint},\eqref{eq:betaAndc}. Applying \eqref{eq:gieselInequality}, \eqref{eq:myDv}, and the fact that $V(x)\geq0$ is affine on each $\sigma_i$ shows that $D^+V(x) \leq \sum_{j=0}^n \alpha_j D^+_{i,j} V \leq -b_2 \sum_{j=0}^n \alpha_j V_{x_{i,j}} = -b_2 V(x)$, where $x=\sum_{j=0}^n\alpha_jx_{i,j}\in\T\degree$, $\sum_{j=0}^n \alpha_j=1$, and $0\leq\alpha_j\leq1$. Like \cite{Doban}, as a relaxation of Theorem\;\ref{thm:gieseldWBound}, it is assumed that $g_{\boldsymbol{\lambda}}(\cdot)\in\mathbb{C}^2(\T)$, not everywhere. Since $x\in\T\degree$ was an arbitrary point, \eqref{eq:myExpoDecrease} is verified.
\end{proof}

Even if $b_2^\ast \leq 0$ in \eqref{eq:genOpt}, a stabilizing controller can be found if $D^+_{i,j}V$ in \eqref{eq:myDv} is positive in all simplexes that include the origin because in this case, a set $\A\in\mathfrak{R}$ can be obtained. This is described in the following. 

\begin{corollary} \label{cor:mightFindAController}
Suppose that $b_2^\ast\leq0$ in \eqref{eq:genOpt}. Let $\mathbb{I}_0=\Set{i\in\IntSet_1^{\mt} | 0\in\sigma_i}$ and $\mathcal{E}_0=\{\sigma_i\}_{i\in\mathbb{I}}$. Further, let $\mathbb{I}_1= \{i\in\IntSet_1^{\mt} \,\mid\, D^+V^\ast_{x_{i,j}} < 0, \;\forall j\in\IntSet_0^n, \; \textrm{ where } x_{i,j}\neq0 \}$, and $\mathcal{E}_1=\{\sigma_i\}_{i\in\mathbb{I}_1}$. If $\mathcal{E}_1\supseteq\mathcal{E}_0$, then $V^\ast:\hat{\mathcal{E}}_1\rightarrow\mathbb{R}$ constructed from the elements of $\textbf{V}^\ast$ is a Lyapunov function of $\dot{x}=g_{\boldsymbol{\lambda}^\ast}(x)$, where $\hat{\mathcal{E}}_1\subseteq\mathcal{E}_1$ is in $\mathfrak{R}^n$. Let $\A = V^{\ast^{-1}}([0,r])\subseteq\hat{\mathcal{E}}_1$ be in $\mathfrak{R}^n$ for some $r>0$. Then $x=0$ is locally exponentially stable for $\dot{x}=g_{\boldsymbol{\lambda}^\ast}(x)$ with $||x(t)||\leq\sqrt[^{a^\ast}]{r/b_1^\ast}e^{-(\hat{b}_2^\ast/a^\ast)(t-t_0)}$ if $x(t_0)\in\A\degree$, where $\hat{b}_2^\ast \coloneqq \min \Set{-D^+_{i,j}V^\ast / V^\ast_{x_{i,j}} \mid i\in\mathbb{I}_1, j\in\IntSet_j^n, x_{i,j} \neq 0 }$. \qed
\end{corollary}

\begin{proof}
For all simplexes in $\mathcal{E}_1$, $D_{i,j}^+V^\ast$ is negative except at $0$ where it is zero, making $\hat{b}^\ast_2$ positive.  $\mathcal{E}_1{\supseteq}\mathcal{E}_0$ ensures that $\hat{\mathcal{E}}_1{\in}\mathfrak{R}^n$ exists because $\mathcal{E}_0{\in}\mathfrak{R}^n$. The claim follows from Theorem\;\ref{eq:genOpt} by letting $\T{\coloneqq}\hat{\mathcal{E}}_1$, and $b^\ast_2{\coloneqq}\hat{b}_2^\ast$ in \eqref{eq:genOpt}.
\end{proof}

In practice, it may not be obvious how to apply Theorem\;\ref{thm:genOpt} and Corollary\;\ref{cor:mightFindAController} for control design. For one, finding a control structure in which point-wise feasibility on vertices of a triangulation implies feasibility at all points in the triangulation, is not trivial. Once the control structure is chosen, its first and second derivatives may need to be constrained to compute $\beta_i$ in \eqref{eq:betaAndc}. Moreover, constraints \eqref{eq:myDv} and \eqref{eq:uConstraintGeneral} are nonlinear. Note that searching for a positive $b_2$ in \eqref{eq:genOpt} is important because even if $\mathcal{E}_1\supseteq\mathcal{E}_0$ in Corollary\;\ref{cor:mightFindAController}, the $\hat{b}_2^\ast$ or $\A$ obtained by it might be too small. A practical design for control-affine systems using Theorem\;\ref{thm:genOpt} is discussed next.

\subsection{Design for Control-Affine Systems}
Let the system in \eqref{eq:controlSystem} be control-affine with a polytopic input constraint. If CPA controllers are chosen as class $\mathcal{F}$ in Theorem\;\ref{thm:genOpt}, which means each element of $u$ is a CPA function, and $\beta_i$ is computed using \eqref{eq:betaAndc}, the only remaining nonlinearities in \eqref{eq:genOpt}'s constraints are bilinear terms in \eqref{eq:myDv}. Using a feasible initialization that has $b_2\!\leq\!0$, convex overbounding can be used to iteratively find larger values for $b_2$ on a fixed triangulation in a process inspired by \cite{warner2017}. This is formulated as an iterative SDP here. The following theorem integrates the computation of $\beta_i$ with \eqref{eq:genOpt} when a CPA controller is sought, and highlights the remaining nonlinearities.

\begin{theorem} \label{thm:semiProg}
Consider the constrained control system
\begin{align} \label{eq:controlAffineSystem}
    \dot{x} = f(x) + G(x)u, \;\;x\in\X\in\mathfrak{R}^n, \;\;u\in\mathcal{U}\in\mathfrak{R}^m,
\end{align}
where $f(0)=0$, and $\mathcal{U}=\Set{u\in\R^m \mid H u \leq h_c}$. Given a triangulation $\T =\{\sigma_i\}_{i=1}^{\mt}$, where $\T\subseteq\X$, suppose that both $f(\cdot),G(\cdot)\in\mathbb{C}^2(\T)$. Let $u$ be CPA on $\T$, i.e. $u^{(s)}:\T \rightarrow \R$, $\forall s\in\IntSet_1^m$, where  ${u^{(s)}}_i = x^\intercal \nabla{u^{(s)}}_i+\omega^{(s)}_i$, $\forall s\in\IntSet_1^m, \forall i\in\IntSet_1^{\mt}$. Let $\textbf{y} =  [\mathbf{V}, \mathbf{L}, \mathbf{U}, \mathbf{Z}, a, \mathbf{b}]$ be the unknowns, where $\mathbf{V}=\{V_x\}_{x\in\Et}\subset\R^n$ and $\mathbf{L}=\{l_i\}_{i=1}^{\mt}\subset\R^n$, and $\mathbf{U}=\{u_x\}_{x\in\Et}\subset\R^m$, and $\mathbf{Z}=\{z_i\}_{i=1}^{\mt}\subset\R$, and $a\in\R$, and  $\mathbf{b}=\{b_1,b_2\}\subset\R$. The following optimization is feasible.
\begin{subequations} \label{eq:SemiProg} 
    \begin{alignat}{2}
        \textbf{y}^\ast &= \argmin_{\textbf{y}} \;\; J(\textbf{y}) \nonumber \\
        \textrm{s.t.} \;\; & V_0 = 0, \;\; a, b_1 > 0, && \label{eq:SemiV0b1Constraint} \\
        & b_1||x||^a \leq V_x, && \forall x\in\Et\backslash\{0\}, \label{eq:SemiVconstraint} \\
        & |{\nabla V}_i| \leq l_i, && \forall i\in\IntSet_1^{\mt}, \label{eq:SemiNablaConstraint} \\
        & u_0 = 0, \; H u_x \leq h_c, && \forall x \in \Et\backslash\{0\}, \label{eq:SemUconstraint} \\
        & |\nabla{u^{(s)}}_i| \leq z_i, && \forall i\in\IntSet_1^{\mt}, \; \forall s\in\IntSet_1^m, \label{eq:SemDuConstraint} \\
        & D^+_{i,j}V \leq -b_2 V_{x_{i,j}}, \quad && \forall i\in\IntSet_1^{\mt}, \; \forall j\in\IntSet_0^n,  \label{eq:semDv} 
        \end{alignat}
\end{subequations}

\noindent where $D^+_{i,j}V=\phi_{i,j} + u_{x_{i,j}}^\intercal G(x_{i,j})^\intercal\nabla{V}_i + c_{i,j}\eta_iz_i1_n^\intercal l_i$ and $\phi_{i,j} = f(x_{i,j})^\intercal {\nabla V}_i {+} c_{i,j}\mu_i 1_n^\intercal l_i$, and $c_{i,j}$ is given in \eqref{eq:betaAndc}, and
\begin{flalign}\label{eq:affineB}
    &\mu_i{=}\max_{p,q,r\in\IntSet_1^n} \max_{\xi\in\sigma_i} \left| \left. \sfrac{\partial^2 f^{(p)}}{\partial x^{(q)}\partial x^{(r)}}\right|_{x=\xi}\right| + \ldots & \\ 
    & \sum_{s=1}^m \left| \left. \sfrac{\partial^2 G^{(p,s)}}{\partial x^{(q)}\partial x^{(r)}} \right|_{x=\xi} \right| \max_{u^{(s)}} \left|\textrm{proj}_s(\mathcal{U})\right|, \textrm{ and } & \nonumber \\
    &\eta_i{=}\!\max_{p,q,r{\in}\IntSet_1^n} \max_{\xi{\in}\sigma_i}\!\sum_{s{=}1}^m \!\left| \left. \sfrac{\partial G^{(p,s)}}{\partial x^{(q)}}\right|_{x{=}\xi} \right|\! {+} \!\left| \left. \sfrac{\partial G^{(p,s)}}{\partial x^{(r)}}\right|_{x{=}\xi} \right|, \nonumber &
\end{flalign}
\noindent where $\textrm{proj}_s(\mathcal{U})$ projects $\mathcal{U}$ onto the $s$-th axis of $\R^m$. \qed
\end{theorem}

\begin{proof}
Consider any $\sigma_i{\in}\T$. By generalizing \cite[Lem\;III.1]{Doban} to multi-input systems with polytopic input constraints, the right-hand side of \eqref{eq:betaAndc} can be bounded above by $\mu_i+\eta_i z_i$, where $z_i\geq |\nabla{u^{(s)}}_i|$, using the Triangle Inequality. Considering $z_i$ as an optimization variable, and replacing $\beta_i$ with $\mu_i{+}\eta_i z_i$, and including $z_i{\geq} |\nabla{u^{(s)}}_i|$ in \eqref{eq:genOpt}, \eqref{eq:SemiProg} is obtained. Thus, the claim follows from Theorem\;\ref{thm:genOpt}.
\end{proof}

The only remaining nonlinearities in \eqref{eq:SemiProg}'s constraints are $||x||^a$ terms in \eqref{eq:SemiVconstraint}, and the bilinear terms in $D^+_{i,j}V{-}\phi_{i,j}$ and the right-hand side of \eqref{eq:semDv}, since $\mu_{i,j},\eta_{i,j}$'s are known constants on a given triangulation, $\T$. Note that there are $(n{+}1)\mt$ constraints in the form of \eqref{eq:semDv}, which grows linearly with $\mt$. Thus, covexifying \eqref{eq:SemiProg} is valuable to make it more practical. Theorem\;\ref{thm:semiProg} can be used as a nonlinear optimization to find a CPA controller, or used to find an initialization for the following iterative SDP algorithm.

\subsection{Iterative Design Algorithm}
The primary objective in both nonlinear optimizations \eqref{eq:genOpt} and \eqref{eq:SemiProg} is finding a $b_2^\ast>0$ to ensure stability. Choosing a cost function that weighs increasing $b_2$ against performance is a bad choice because no controller is formulated until  $b_2>0$ is found. This section gives an algorithm for system \eqref{eq:controlAffineSystem} that iteratively searches for $b_2>0$ using a sequence of SDPs. If a sufficiently large $b_2>0$ is found, the algorithm fixes it, and then optimizes other performance objectives in another sequence. The following theorem formulates each iteration.

\begin{theorem} \label{thm:semiProg2}
Suppose that $J$ in \eqref{eq:SemiProg} is linear or quadratic, and $a>0$ is a fixed number. Let $\underline{\mathbf{y}}=[\underline{\mathbf{V}}, \underline{\mathbf{L}}, \underline{\mathbf{U}}, \underline{\mathbf{Z}}, a,  \underline{\mathbf{b}}]$ satisfy \eqref{eq:SemiVconstraint}--\eqref{eq:semDv}. Consider the following optimization.
\begin{subequations} \label{eq:SemiProg2} 
    \begin{alignat}{2}
        &\delta{\textbf{y}^\ast} = \argmin_{\delta\mathbf{y} =  [\delta\mathbf{V}, \delta\mathbf{L}, \delta\mathbf{U}, \delta\mathbf{Z}, 0, \delta\mathbf{b}]} && J(\underline{\textbf{y}}+\delta\textbf{y}) \nonumber \\
        &\textrm{s.t.} && \nonumber \\
        & \delta V_0 = 0, \;\; \underline{b}_1 + \delta b_1 > 0, && \label{eq:Semi2V1b0Constraint} \\
        & (\underline{b}_1+\delta b_1)||x||^a \leq \underline{V}_x +  \delta V_x, && \forall x\in\Et\backslash\{0\}, \label{eq:SemiVconstraint2} \\
        & |{\nabla\underline{V}}_i + \delta {\nabla V}_i| \leq \underline{l}_i + \delta l_i, \quad && \forall i\in\IntSet_1^{\mt},  \label{eq:SemiNablaConstraint2} \\
        & \delta u_0 = 0, \; H (\underline{u}_x + \delta u_x) \leq h_c, && \forall x \in \Et, \label{eq:SemUconstraint2} \\
        &|\nabla{\underline{u}^{(s)}}_i {+} \delta \nabla{u^{(s)}}_i| \leq \underline{z}_i {+} \delta z_i, \;\; && \forall i\in\IntSet_1^{\mt}, \; \forall s\in\IntSet_1^m, \label{eq:SemDuConstraint2} \\
        & P_{i,j} \leq 0, && \forall i\in\IntSet_1^{\mt}, \; j{=}0, \label{eq:2SemDv1} \\
        & Q_{i,j} \leq 0, && \forall i\in\IntSet_1^{\mt}, \; \forall j\in\IntSet_1^n, \label{eq:2SemDv2}
        \end{alignat}
\end{subequations}

\noindent where $\delta{\nabla V}_i{=}X_i^{-1}\delta\bar{V}_i$, $\delta{\nabla u^{(s)}}_i{=}X_i^{-1}\delta\bar{u}_i$ as in Remark\;\ref{rem:nablaLinear},
\begin{equation}
P_{i,j} = \begin{bmatrix} \hat{\phi}_{i,j} & \ast & \ast & \ast & \ast \\
                          \delta{\nabla{V}}_i & -2I_n & \ast & \ast & \ast \\
                          G(x_{i,j})\delta u_{x_{i,j}} & 0 & -2I_n & \ast & \ast \\
                          \delta V_{x_{i,j}} & 0 & 0 & -2 & \ast \\
                          \delta b_2 & 0 & 0 & 0 & -2 \end{bmatrix},
\end{equation}

\begin{align}
    \hat{\phi}_{i,j} = & ({\nabla \underline{V}}_i + \delta{\nabla V}_i)^\intercal(f(x_{i,j}) + G(x_{i,j}) \underline{u}_{x_{i,j}}) + \ldots \nonumber \\
    & {\nabla \underline{V}}_i^\intercal G(x_{i,j})\delta u_{x_{i,j}} + \mu_ic_{i,j}1_n^\intercal (\underline{l}_i+\delta l_i)  + \ldots \nonumber \\
    & \eta_ic_{i,j}\left(( \underline{z}_i+\delta z_i)1_n^\intercal  \underline{l}_i + \underline{z}_i 1_n^\intercal\delta l_i \right) + \ldots  \nonumber \\
    & b_2 ( \underline{V}_{x_{i,j}}  + \delta V_{x_{i,j}}) +  \underline{V}_{x_{i,j}}\delta b_2, \textrm{ and} \label{eq:phiHat} 
\end{align}

\begin{equation}
Q_{i,j} = \begin{bmatrix} P_{i,j} & \ast & \ast\\
                          1_n^\intercal\delta l_i & \frac{-2}{\eta_i c_{i,j}} & \ast \\
                          \delta z_i & 0 & \frac{-2}{\eta_i c_{i,j}} \end{bmatrix},
\end{equation}

\noindent and $c_{i,j}$ is given in \eqref{eq:betaAndc}, and $\mu_i, \eta_i$ are given in \eqref{eq:affineB}. Then, $\underline{\textbf{y}}+\delta\textbf{y}^\ast$ is a feasible point for \eqref{eq:SemiProg}, and $J(\underline{\textbf{y}}+\delta\textbf{y}^\ast)\leq J(\underline{\textbf{y}})$. \qed
\end{theorem}

\begin{proof}
To see that \eqref{eq:SemiProg2} is feasible, observe that $\delta\textbf{y}{=}0$ satisfies \eqref{eq:SemiProg2} since in this case, \eqref{eq:SemiProg2} is equivalent to \eqref{eq:SemiProg} with $\textbf{y}:=\underline{\textbf{y}}$. In fact, \eqref{eq:2SemDv1}--\eqref{eq:2SemDv2} are the convexified equivalences of \eqref{eq:semDv}. To show this, recall that $w^\intercal v {\leq} 1/2(w^\intercal w {+}v^\intercal v)$ for any $v,w$ vectors with the same dimension. Applying this fact with $(v,w){=}(\delta \nabla{V}_i, G(x_{i,j}) \delta u_{x_{i,j}})$, $(v,w){=}( \delta z_i, 1_n^\intercal\delta l_i)$, and $(v,w){=}(\delta V_{x_{i,j}}, \delta b_2)$ shows that by Schur complement, \eqref{eq:2SemDv1} is implied when $j{=}0$, since $c_{i,j}$ is zero in this case, and \eqref{eq:2SemDv2} is implied when $j{\neq} 0$. Finally, $J(\underline{\mathbf{y}} {+} \delta \mathbf{y}) \leq J(\mathbf{y})$ because otherwise $\delta\mathbf{y}=0$ would be a better, feasible solution.
\end{proof}

\begin{remark}
In case $G(x)$ in \eqref{eq:controlAffineSystem} is a constant matrix, \eqref{eq:2SemDv1} must be used for all $j\in\IntSet_0^n$ because in this case, $\eta_i=0$ in \eqref{eq:affineB}. Also, \eqref{eq:SemDuConstraint} and \eqref{eq:SemDuConstraint2} are not needed. \qed
\end{remark}

Starting with a feasible point of \eqref{eq:SemiProg},  Theorem\;\ref{thm:semiProg2} can be used repeatedly to potentially decrease the values of the cost function. Note that by replacing $\delta V_0=\delta u_0=0$ in the simplexes that have $0$ as their vertex, and letting $\underline{b}_1+\delta b_1$ be greater than or equal to a small positive number, \eqref{eq:SemiProg2} is a SDP in the standard format. The small positive number must be kept constant in the later iterations. Two methods of finding a feasible initialization point are given next.

\begin{initialization} \label{initialization:random}
Choosing $a,b_1>0$, let $V_x=b_1||x||^a$, $\forall x{\in}\Et$. Let $u_0{=}0$ and assign admissible $u_x$ for all $x{\in}\Et$. They can be random. Compute $l_i{=}|\nabla{V}_i|$ and $z_i{=}|\nabla{u^{(s)}}_i|$ for all $i\in\IntSet_1^{\mt}$ as in Remark\;\ref{rem:nablaLinear}. Finally, find the largest $b_2$ satisfying \eqref{eq:semDv} in all simplexes. \qed
\end{initialization}

\begin{initialization} \label{initialization:LQR}
Linearize \eqref{eq:controlAffineSystem} around the origin. Design a LQR controller, and find the corresponding quadratic Lyapunov function, $x^\intercal\hat{P}x$. Sample $x^\intercal\hat{P}x$ at the vertices of $\T$ to find $\mathbf{V}$, and let $a{=}2$ and $b_1$ be equal to the smallest eigenvalue of $\hat{P}$. Sample the LQR controller at the vertices of $\T$ to form $\mathbf{U}^{\textrm{LQR}}{=}\{u_x^{\textrm{LQR}}\}_{x\in\Et}$. Divide each element of $\mathbf{U}^{\textrm{LQR}}$ by a positive number so that the result, $\mathbf{U}{=}\{u_x\}_{x\in\Et}$, has admissible values for all vertices. Compute $l_i{=}|\nabla{V}_i|$ and $z_i{=}|\nabla{u^{(s)}}_i|$ for all $i{\in}\IntSet_1^{\mt}$ as in Remark\;\ref{rem:nablaLinear} using the computed values of $V_x$ and $u_x$, respectively. Finally, find the largest $b_2$ satisfying \eqref{eq:semDv} in all simplexes. \qed
\end{initialization}

Given a triangulation and a linear or quadratic cost function $\hat{J}(\textbf{V},\textbf{L},\textbf{U},\textbf{Z},b_1)$, the procedure for finding a stabilizing CPA controller for \eqref{eq:controlAffineSystem} is given in Algorithm\;\ref{alg:cpaControl}. It iteratively increases $b_2$ until it is positive. Since $e^{-(\sfrac{b_2}{a})t}$ is proportional to the state norm's upper-bound when $a{>}0$ is fixed, increasing $b_2{>}0$ can continue until a desired decay rate is ensured. Then, by fixing $b_2$'s value, $\hat{J}(\cdot)$ is iteratively minimized. Finally, the corresponding positive-invariant set, $\A{=}{V}^{-1}([0,r])$, $r{>}0$, where $\A{\subseteq}\T$ and $\A{\in}\mathfrak{R}^n$, is found. Both of the loops can be terminated in lines \ref{line:term1} and \ref{line:term2} if a predefined maximum number of iterations is reached. If a sufficiently large positive $b_2$ cannot be found, triangulation refinement, discussed later, is needed. 

\begin{algorithm}[h!]
	\caption{CPA control design on a fixed triangulation}
    \label{alg:cpaControl}
	\begin{algorithmic}[1]
	   \Require The control-affine system \eqref{eq:controlAffineSystem}, and a triangulation $\T\subseteq\X$, and a linear or quadratic $\hat{J}(\textbf{V},\textbf{L},\textbf{U},\textbf{Z},b_1)$ 
	   \Ensure $u(x)$, and a positive-invariant set $\A$
	   \State $\underline{\mathbf{y}} \coloneqq $ a feasible point of \eqref{eq:SemiProg} (using Initialization\;\ref{initialization:random} or \ref{initialization:LQR})
	   \State $J \coloneqq -b_2$ \Comment{since $b_2$ is to be maximized}
	   \Repeat{}
	       \State Use Theorem\;\ref{thm:semiProg2}
	   \Until{$b_2 > 0$ is large enough OR $b_2$ is not changing} \label{line:term1}
	   \If{$b_2>0$ is found}
	       \State Fix $b_2$, and let $J\coloneqq\hat{J}(\cdot)$
	       \Repeat{}
	           \State Use Theorem\;\ref{thm:semiProg2}
	       \Until{$J$ is sufficiently small OR $J$ is not changing} \label{line:term2}
	       \State Return $u(x)$ and find a set $\A={V}^{-1}([0,r])$, $r>0$, \\ \hspace{0.4cm} where $\A\subseteq\T$ and $\A\in\mathfrak{R}^n$ \label{line:findingA}
	   \EndIf
	\end{algorithmic}
\end{algorithm}

Once Algorithm\;\ref{alg:cpaControl} returns the stabilizing controller, the corresponding point, $\mathbf{y}$, can serve as the initial guess for the non-convex optimization \eqref{eq:SemiProg} with a nonlinear cost function $J(\cdot)$, as a final attempt to boost the performance.

\subsection{Minimum-norm Online Implementation}
Consider system \eqref{eq:controlAffineSystem}. Algorithm\;\ref{alg:cpaControl} only minimizes the objective pointwise on the vertices of the triangulation unless $\hat{J}(\cdot)$ is chosen wisely. However, since the corresponding Lyapunov function of the returned controller is also a Lipschitz CLF, a minimum-norm controller can be formulated as a QP \cite{Ames2016,nonMPC-Lyap2006}. Suppose that $b_2^\ast>0$ is found by Algorithm\;\ref{alg:modification}, and $V^\ast$ is the corresponding CPA Lyapunov function. Let $\A$ be $\A{=}{V^\ast}^{-1}([0,r])$, $r{>}0$, where $\A{\subseteq}\T$ and $\A{\in}\mathfrak{R}^n$. Starting at any $x{\in}\A\degree$, the minimum-norm controller can be written as 
\begin{subequations} \label{eq:minNormLips} 
    \begin{flalign} 
        &u^\ast(x) = \argmin_{u} \;\; u^\intercal\hat{H}(x)u + \hat{h}(x)^\intercal u & \nonumber \\
        &\textrm{s.t.\;\;} Hu\leq h_c, & \\
        & \quad\;\;\, \nabla{V^\ast_i}^\intercal(f(x){+}G(x)u){+} b_2^\ast V^\ast(x) \leq 0, \;\; \forall i\in\mathcal{I}, &
        \end{flalign}
\end{subequations}

\noindent where $\mathcal{I} {=} \Set{i{\in}\IntSet_1^{\mt} | x{\in}\sigma_i}$, and $\hat{H}(x)$ is positive definite. The set $\mathcal{I}$ has more than one element if $x$ is on the common face of some simplexes. The optimization \eqref{eq:minNormLips} is feasible for all $x\in\A$, because the corresponding CPA controller of $V^\ast$ is a feasible point for it. Therefore, the convergence inequality $||x(t)||\leq\sqrt[^a]{r/b_1}e^{-(b_2/a)(t-t_0)}$ that holds for the CPA controller, also holds for the QP-based controller. 

\section{Triangulation Refinement}
Both Theorem\;\ref{thm:genOpt} and Algorithm\;\ref{alg:cpaControl} work on given fixed triangulations. If a positive $b_2$ cannot be found, the triangulation can be refined. These refinements can be local by tracking the value of $D^+_{i,j}V$ on the simplexes in $\T$. However, for simplicity, a structured triangulation with uniform refinement over all simplexes is proposed here.

The standard triangulation, denoted by $\T^{\textrm{std}}$, is a hyper-cube in $\R^n$ composed of generalized isosceles right triangle simplexes with unit length sides that can be tessellated to cover the whole $\R^n$ \cite[Sec\;3.1]{gieslRevCPA2013}. Multiplying all of its vertices with a positive number scales the triangulation.

\begin{definition} \label{def:scaledSubset}
Given $\X\in\mathfrak{R}^n$ and $\rho>0$, a scaled subset of the $\T^{\textrm{std}}$ in $\X$, denoted $\T_\X^\rho$, is a triangulation obtained by scaling $\T^\textrm{std}$ with $\rho$, and then finding the largest collection of its simplexes entirely in $\X$. \qed
\end{definition}

Let the volume enclosed by $\Omega\in\mathfrak{R}^n$ be denoted by $\textrm{vol}(\Omega)$. Given an initial $\rho$, and a minimum threshold on the covering percentage of $\X$, denoted by $\epsilon_c$, Algorithm\;\ref{alg:modification} finds a small enough $\rho$ so that $\textrm{vol}(\T_\X^h)/\textrm{vol}(\X)\geq \epsilon_c$. Then it searches for a controller. If not successful, $\rho$ is decreased to refine all simplexes, and the search continues. The algorithm terminates if $u(x)$ is returned, or if decreasing $\rho$ finally violates a given threshold, $\rho_{\textrm{min}}$.

\begin{algorithm}[h!]
	\caption{Control design with triangulation refinement}
    \label{alg:modification}
	\begin{algorithmic}[1]
	   \Require System \eqref{eq:controlSystem} (or \eqref{eq:controlAffineSystem}), cost function, $\rho$, $0<\gamma<1$, $\epsilon_c$.
	   \Ensure $u(x)$ and a positive-invariant set $\A$
	   \Repeat{}
	        \State $\T \coloneqq \T_\X^\rho$, where $\textrm{vol}(\T_\X^\rho)/\textrm{vol}(\X)\geq \epsilon_c$
	        \State Solve \eqref{eq:genOpt} (or  Algorithm\;\ref{alg:cpaControl})
	        \If{$b_2>0$ is found}
	            \State Return: output of Theorem\;\ref{thm:genOpt} (or  Algorithm\;\ref{alg:cpaControl}) \label{line:refFindingA}
	        \EndIf
	        \State $\rho := \gamma \rho$ and make sure $\textrm{vol}(\T_\X^\rho)/\textrm{vol}(\X)\geq \epsilon_c$
	   \Until{$\rho < \rho_{\textrm{min}}$ OR $b_2>0$ }
	\end{algorithmic}
\end{algorithm}

\begin{remark}
Since \eqref{eq:SemiProg2}'s solution satisfies \eqref{eq:genOpt}, Whenever \eqref{eq:genOpt} is solved or Theorem\;\ref{thm:semiProg2} in Algorithm\;\ref{alg:cpaControl} is invoked, and a $b_2{\leq}0$ is found, Corollary\;\ref{cor:mightFindAController} can be checked to see if a positive $\hat{b}_2$ exists. If so, $\A$ is also obtained by Corollary\;\ref{cor:mightFindAController}. \qed
\end{remark}
\section{Numerical Simulation}
\begin{figure}
    \centering
    \includegraphics[width=8.65cm]{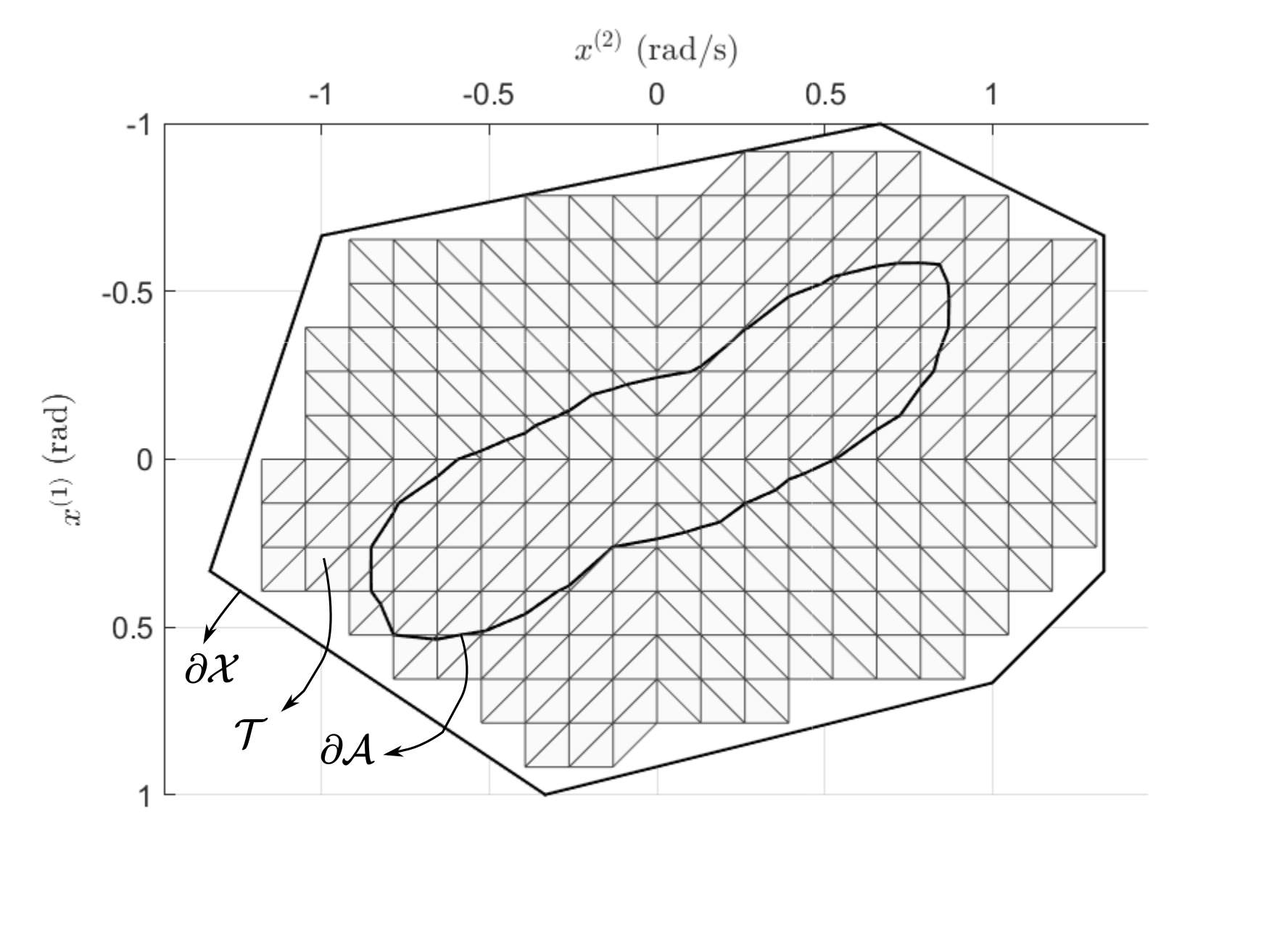}
    \caption{The set $\X$, the triangulation $\T\subseteq\X$, and $\partial\A$, the boundary of a sub-level set of the CPA Lyapunov function in $\T$.}
    \label{fig:trian}
\end{figure}

\begin{figure}
    \centering
    \includegraphics[width=8.65cm]{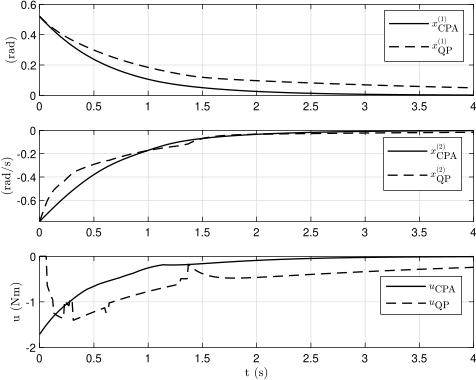}
    \caption{State and input trajectories of both controllers. Since $u^\ast(x)$ in \eqref{eq:minNormLips} may not be Lipschitz \cite{Ames2016}, occasional jumps in $u_{\textrm{QP}}$ happen.}
    \label{fig:traj}
\end{figure}

Consider the inverted pendulum $\dot{x}^{(1)} = x^{(2)}$, $\dot{x}^{(2)} = 4.9\sin{x^{(1)}} - 0.3x^{(2)} + u$, where the polytope $\X$ in Fig.\ref{fig:trian} and $|u|\leq 5$ define its state and input constraints. All units are SI. To solve SDPs, Yalmip \cite{yalmip} with Sedumi \cite{sedumi} were used in MATLAB. For initialization, an LQR with the cost function $x^TQx+u^2$, where $Q=2I$, was used. Choosing $\rho=0.5$, $\gamma = 0.8$, and $\epsilon_c = 0.85$ in Algorithm\;\ref{alg:modification}, and limiting the convex-overbounding iterations in Algorithm\;\ref{alg:cpaControl} to five, a CPA controller was found on $\T_\X^{0.13}$ with $b_2=0.33$ and $b_1=4.27$ after five iterations. Corollary\;\ref{cor:mightFindAController} could be used in the second iteration to return a solution earlier, however, refining the triangulation resulted in a larger positive invariant set, $\A$. The triangulation and the boundary of the Lyapunov function's sub-level set $\A\subseteq\T$ satisfying $\A\in\mathfrak{R}^n$ are depicted in Fig.\;\ref{fig:trian}. No further improvement was made offline. To simulate the QP-based controller \eqref{eq:minNormLips}, the objective function $u^2$ was used in \eqref{eq:minNormLips}, and the system equations were integrated using the 4th order Runge-Kutta with a $0.01$\;s time step. Starting at $x=(0.52,-0.78)$, which is inside the level set, the state trajectories and inputs of the two controllers are given in Fig.\;\ref{fig:traj}. Since the state has a different evolution using the QP-based controller, the trajectory of its input, $u_{\textrm{QP}}$, is not always below the trajectory of the CPA controller input, $u_\textrm{CPA}$, despite the fact that $u_{\textrm{CPA}}$ is a minimum-norm realization. The time it takes for both states to settle in $\pm0.05$ range for the CPA controller is $1.8$\;s and is $4.1$\;s for the QP-based controller. Since the initial point is inside the sub-level set, the state and input constraints are respected.

\section{Conclusion}
In this paper, a method to stabilize state and input constrained nonlinear systems was proposed via an offline optimization on a variable triangulation. The method provides an exact region of attraction, and bounds the decay rate of the state norm above. For control-affine systems, the optimization was formulated by iterative SDPs assuming CPA structure for the controller. In this case, the corresponding Lyapunov function, which is a Lipschitz CLF, was used to also formulate a minimum-norm QP-based controller. 
\bibliographystyle{unsrt}  
\bibliography{BibliographyShort.bib}

\begin{thebibliography}{10}

\bibitem{Ames2016}
A.~D Ames, X.~Xu, J.~W Grizzle, and P.~Tabuada.
\newblock Control barrier function based quadratic programs for safety critical
  systems.
\newblock {\em IEEE Trans. Aut. Ctrl}, 62(8):3861--3876, 2016.

\bibitem{robot2020}
J.~Nubert, J.~Köhler, V.~Berenz, F.~Allgöwer, and S.~Trimpe.
\newblock Safe and fast tracking on a robot manipulator: Robust {MPC} and
  neural network control.
\newblock {\em IEEE Robot. Aut. Letters}, 5(2):3050--3057, 2020.

\bibitem{chemProccess2013}
R.~Amrit, J.~B. Rawlings, and L.~T. Biegler.
\newblock Optimizing process economics online using model predictive control.
\newblock {\em Computers \& Chem. Eng.}, 58:334--343, 2013.

\bibitem{nonMPC1998}
H~Chen and F~Allg{\"o}wer.
\newblock Nonlinear model predictive control schemes with guaranteed stability.
\newblock In {\em Nonlinear model based process control}, pages 465--494.
  Springer, 1998.

\bibitem{nonMPCeconomic2018}
T.~Faulwasser, L.~Gr{\"u}ne, M.~A M{\"u}ller, et~al.
\newblock {\em Economic nonlinear model predictive control}.
\newblock Now Foundations and Trends, 2018.

\bibitem{nonMPC-Lyap2005}
A.~Jadbabaie and J.~Hauser.
\newblock On the stability of receding horizon control with a general terminal
  cost.
\newblock {\em IEEE Trans. Aut. Ctrl}, 50(5):674--678, 2005.

\bibitem{nonMPC-Lyap2006}
P.~Mhaskar, N.~H El-Farra, and P.~D Christofides.
\newblock Stabilization of nonlinear systems with state and control constraints
  using {L}yapunov-based predictive control.
\newblock {\em Sys. Ctrl Letters}, 55(8):650--659, 2006.

\bibitem{nonMPCefficient2009}
M.~Diehl, H.~J. Ferreau, and N.~Haverbeke.
\newblock Efficient numerical methods for nonlinear {MPC} and moving horizon
  estimation.
\newblock In {\em Nonlinear model predictive control}, pages 391--417.
  Springer, 2009.

\bibitem{nonMPCefficient2020}
S.~Gros, M.~Zanon, R.~Quirynen, A.~Bemporad, and M.~Diehl.
\newblock From linear to nonlinear {MPC}: bridging the gap via the real-time
  iteration.
\newblock {\em Int. J. Ctrl}, 93(1):62--80, 2020.

\bibitem{nonMPCExTube2006}
F.~A Bayer, F.~D Brunner, M.~Lazar, M.~Wijnand, and F.~Allg{\"o}wer.
\newblock A tube-based approach to nonlinear explicit {MPC}.
\newblock In {\em Conf. Decision and Ctrl}, pages 4059--4064. IEEE, 2016.

\bibitem{nonMPCexplicit2012}
Alexandra Grancharova and Tor~Arne Johansen.
\newblock {\em Explicit nonlinear model predictive control: Theory and
  applications}, volume 429.
\newblock Springer Science \& Business Media, 2012.

\bibitem{CBF2007}
P.~Wieland and F.~Allg{\"o}wer.
\newblock Constructive safety using control barrier functions.
\newblock {\em IFAC Proc. Vols}, 40(12):462--467, 2007.

\bibitem{CBF2016}
M.~Z. Romdlony and B.~Jayawardhana.
\newblock Stabilization with guaranteed safety using control
  {L}yapunov--barrier function.
\newblock {\em Aut.}, 66:39--47, 2016.

\bibitem{CBF2018}
J{\'A}~Acosta, A.~D{\`o}ria-Cerezo, and E~Fossas.
\newblock Stabilisation of state-and-input constrained nonlinear systems via
  diffeomorphisms: A {S}ontag's formula approach with an actual application.
\newblock {\em Int. J. Robust Nonlin. Ctrl}, 28(13):4032--4044, 2018.

\bibitem{uConCLF1995}
Y.~Lin and E.~D Sontag.
\newblock Control-{L}yapunov universal formulas for restricted inputs.
\newblock {\em Ctrl Theory Adv. Tech.}, 10(4), 1995.

\bibitem{ExpMPCLike2017}
I.~Pejcic, M.~Korda, and C.~N Jones.
\newblock Control of nonlinear systems with explicit-{MPC}-like controllers.
\newblock In {\em Conf. Decision Ctrl}, pages 4970--4975. IEEE, 2017.

\bibitem{gieslRevCPA2013}
P.~A Giesl and S.~F Hafstein.
\newblock Revised {CPA} method to compute {L}yapunov functions for nonlinear
  systems.
\newblock {\em J. Math. Analysis and Apps}, 410(1):292--306, 2014.

\bibitem{Doban}
T.~RV Steentjes, A.~I Doban, and M.~Lazar.
\newblock Feedback stabilization of nonlinear systems:“universal”
  constructions towards real-life applications.
\newblock Master's Thesis, Eindhoven University of Technology, 2016.

\bibitem{khalil}
H.K. Khalil.
\newblock {\em Nonlinear Systems}.
\newblock Pearson Edu. Prentice Hall, 2002.

\bibitem{warner2017}
EC~Warner and JT~Scruggs.
\newblock Iterative convex overbounding algorithms for {BMI} optimization
  problems.
\newblock {\em IFAC}, 50(1):10449--10455, 2017.

\bibitem{yalmip}
J.~{Lofberg}.
\newblock Yalmip : a toolbox for modeling and optimization in {MATLAB}.
\newblock In {\em IEEE Int. Conf. Robot. Aut.}, pages 284--289, Sep. 2004.

\bibitem{sedumi}
J.~F. Sturm.
\newblock Using {S}e{D}u{M}i 1.02, a {MATLAB} toolbox for optimization over
  symmetric cones.
\newblock {\em Optim. Methods Softw.}, 11(1-4):625--653, 1999.

\end{thebibliography}

\begin{acronym}
\acro{SDP}{semi-definite program}
\acro{MPC}{model predictive control}
\acro{CLF}{control Lyapunov function}
\acro{CPA}{continuous piecewise affine}
\acro{QP}{quadratic programming}
\end{acronym}

\end{document}